\documentclass[english,aps,preprint]{revtex4}
\usepackage{graphicx}
\usepackage{amssymb}

\makeatletter

\usepackage{babel}
\makeatother

\begin{document}

\title{Some comments on embedding inflation in the AdS/CFT correspondence}

\author{David A. Lowe}

\email{lowe@brown.edu}

\affiliation{Physics Department, Brown University, Providence, RI 02912, USA}

\begin{abstract}
The anti-de Sitter space/conformal field theory correspondence (AdS/CFT)
can potentially provide a complete formulation of string theory on
a landscape of stable and metastable vacua that naturally give rise
to eternal inflation. As a model for this process, we consider bubble
solutions with de Sitter interiors, obtained by patching together
dS and Schwarzschild-AdS solutions along a bubble wall. For an interesting
subclass of these solutions the bubble wall reaches spacelike infinity
in the black hole interior. Including the effects of perturbations
leads to a null singularity emanating from this point. Such solutions
are interpreted as states in a single CFT, and are shown to be compatible
with holographic entropy bounds. The construction suggests de Sitter
entropy be interpreted as the total number of degrees of freedom in
effective field theory, with a novel adaptive stepsize cutoff.
\end{abstract}
\maketitle

\section{introduction}

In recent years convincing evidence has accumulated that string theory
has a vast landscape of consistent vacuum states, along with numerous
long-lived metastable states that may well be relevant for realistic
cosmology\citep{Kachru:2003aw}. It is important to develop tools
in string theory that are capable of describing transitions between
such states. Currently the AdS/CFT formulation of string theory \citep{Maldacena:1997re}
is the most promising nonperturbative formulation of string theory.
It is the purpose of the present paper to study how this landscape
might be embedded in this framework, building on the earlier work
of \citep{Alberghi:1999kd,Freivogel:2005qh}.

We begin by reviewing classical bubble solutions in asymptotically
AdS space, following the original work of \citep{Blau:1986cw}. For
many of the most interesting solutions, the bubble wall appears to
reach AdS infinity in finite time. We show this situation is unstable
to perturbations, and that instead a null singularity emanates from
this point. We discuss how this is to be interpreted in terms of the
CFT, and propose a new interpretation of de Sitter entropy compatible
with this picture.

\section{Review of classical bubble solutions}

\begin{figure}
\includegraphics{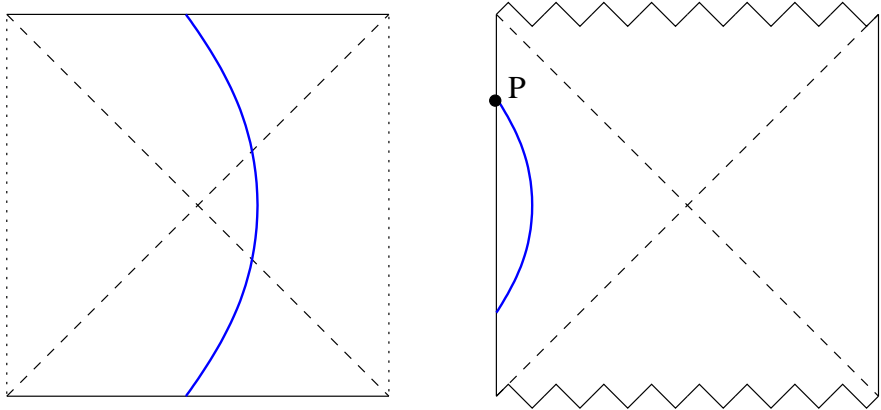}

\caption{Penrose diagrams showing a time symmetric bubble trajectory. The left
diagram refers to de Sitter space, the right diagram refers to Schwarzschild-anti
de Sitter space. The bubble trajectory is shown in blue. The right
segment of Schwarzschild-anti de Sitter corresponds to the exterior
of the bubble. The left segment of de Sitter corresponds to the interior.\label{fig:time symmetric}}

\end{figure}

\begin{figure}
\includegraphics{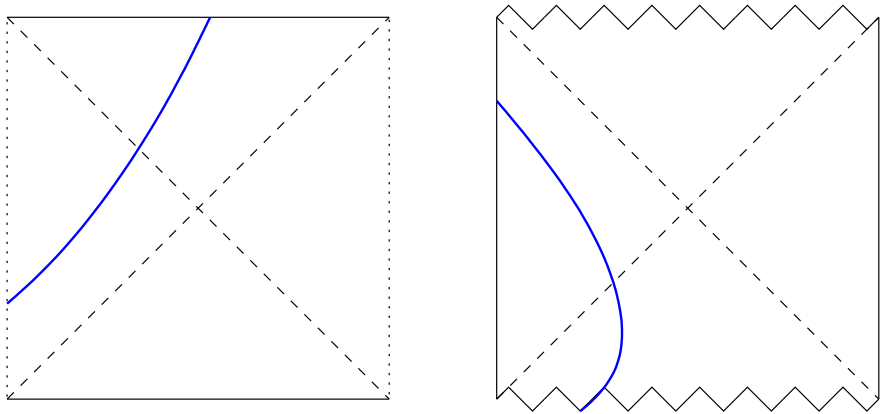}

\caption{Penrose diagrams showing a time asymmetric bubble trajectory.\label{fig:time asymmetric}}

\end{figure}

In order to obtain solutions describing bubbles with inflating interiors
inside asymptotically anti-de Sitter space, we follow the procedure
of Blau, Guendelman and Guth \citep{Blau:1986cw}. The basic idea
is to consider a spherically symmetric bubble in the thin-wall limit.
The interior of the bubble is modeled by a piece of pure de Sitter
space and the exterior by Schwarzschild-anti de Sitter space. This
reduces the problem to specifying the radial position of the bubble
as a function of time which reduces to the one-particle potential
scattering problem described in \citep{Alberghi:1999kd}. In \citep{Alberghi:1999kd}
the case of general dimension and charge was considered. Here we will
specialize to the case of vanishing charge and asymptotically $AdS_{4}$
spacetime. These solutions are similar to those considered in \citep{Blau:1986cw}
and were studied in extensively in \citep{Freivogel:2005qh}. The
main features of all the solutions can be found in the following two
examples in figure \ref{fig:time symmetric} and \ref{fig:time asymmetric},
that we will discuss in detail.%

\begin{figure}
\includegraphics{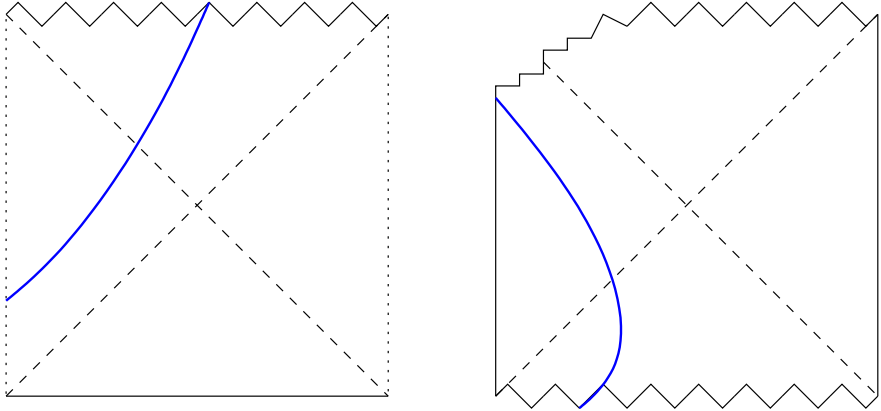}

\caption{Time asymmetric perturbed bubble solution.\label{fig:Time-asymmetric-perturbed}}

\end{figure}

It is important to understand how these solutions change when subject
to small perturbations. One issue is whether the spherical symmetric
approximation is valid. This was considered some time ago in the context
of the original work by Blau et al. \citep{Garriga:1991ts,Garriga:1991tb}.
While perturbations do indeed grow in the de Sitter interior, these
do not qualitatively change the behavior of the solutions. Another
issue peculiar to the situation with negative cosmological constant
is what happens to the asymptotically anti-de Sitter boundaries that
appear on the left. This is investigated in appendix \ref{sec:Appendix:-Generic-perturbations}.
For the time asymmetric case, figure \ref{fig:time asymmetric}, the
conclusion is that a null singularity emerges from the point at which
the bubble wall hits spacelike infinity. When one goes beyond leading
order in perturbation theory, we believe this singularity should become
spacelike, becoming null as it approaches the bubble wall. This is
shown in figure \ref{fig:Time-asymmetric-perturbed}.

Since these solutions are ultimately to be embedded in string theory,
the de Sitter regions will be metastable. We will assume the anti-de
Sitter vacuum under consideration is completely stable. There will
therefore be tunneling between the two solutions. In particular, any
timelike trajectory in the de Sitter interior will tunnel back to
the vicinity of the anti de Sitter vacuum state, with a timescale
exponentially smaller than the Poincare recurrence time of the de
Sitter space. As shown in \citep{Coleman:1980aw,Abbott:1985kr} these
regions undergo a crunch. We have schematically shown this in figure
\ref{fig:Time-asymmetric-perturbed} as a future singularity in the
de Sitter Penrose diagram. 

In fact, the situation is more complicated than this, because while
a given timelike curve will eventually enter a crunching region, the
volume of a given set of compact spacelike slices will increase sufficiently
fast that there is always remnant de Sitter space left at arbitrarily
late time (see for example \citep{Linde:1991sk}). However for the
purpose of predicting experiments measured by a local observer, a
locally defined measure, implicit in figure \ref{fig:Time-asymmetric-perturbed},
is most useful, rather than a volume weighted measure. 

\begin{figure}
\includegraphics{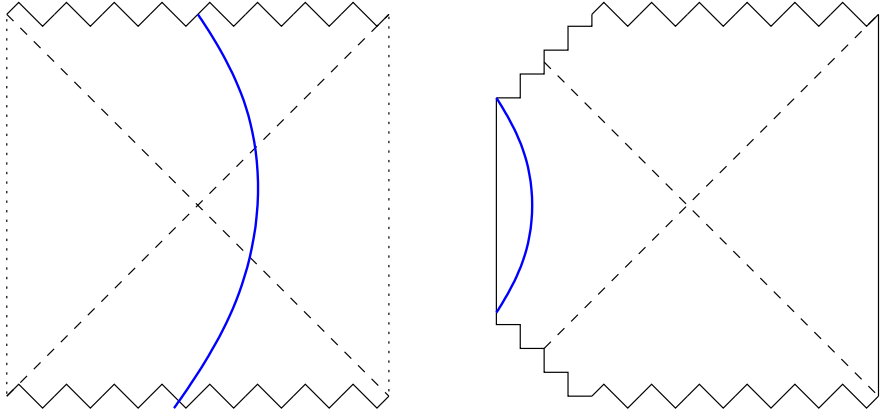}

\caption{Time symmetric perturbed bubble solution.\label{fig:Time-symmetric-perturbed}}

\end{figure}

The same considerations apply to the time symmetric situation of figure
\ref{fig:time symmetric}. Once again, perturbations will generate
a singularity emerging from the endpoint of the bubble wall, extending
into the exterior Schwarzschild-anti de Sitter region. Of course,
if one demands time symmetry, this will also extend outward from the
starting point of the bubble wall. Likewise, on the interior, one
must also take account of the fact that any timelike path will eventually
tunnel to a crunch, so the asymptotic de Sitter future (and past)
is effectively removed. This is shown in figure \ref{fig:Time-symmetric-perturbed}.
Solutions of the type shown in figure \ref{fig:time symmetric} can
only be obtained by a high degree of fine tuning. It seems unlikely
such solutions would emerge from the underlying conformal field theory
description for finite values of Planck's constant.

The version of the AdS/CFT correspondence advocated in \citep{Witten:1998qj}
maps geometries with a fixed boundary conditions at infinity (modulo
conformal transformations) to conformal field theory on the boundary.
We have seen that when perturbations are taken into account the bubble
solutions have only a single asymptotically AdS region, therefore
we expect a description in terms of a single conformal field theory.
The simplest interpretation is in terms of pure states in the CFT
\footnote{Arguments for a mixed state interpretation are presented in \citep{Freivogel:2005qh}.%
}. That said, it is a subtle question how one constructs approximate
local quantities behind the horizon for geometries such as figure
\ref{fig:Time-asymmetric-perturbed}. In the work of \citep{Hamilton:2006fh},
where static black holes were consider, this was accomplished using
analyticity, so that in general CFT correlators needed to be continued
to complex values of space and time. For eternal Schwarzschild-AdS
black holes, this can be reformulated in terms of the thermofield
double formalism, and hence mixed states as in \citep{Freivogel:2005qh}.
However unlike in \citep{Freivogel:2005qh}, we no longer have another
asymptotically AdS exterior region behind the horizon, so it is no
longer possible to construct a precisely defined set of observables
associated with this region. So there is no obvious inconsistency
with a pure state description. We consider the implications of this
picture and the quantum consistency of these solutions, once we have
tackled the issue of interpreting de Sitter entropy in this framework.

\section{de Sitter Entropy}

Each of the bubble solutions considered above share the property that
the expanding de Sitter region is behind a black hole event horizon
from the viewpoint of the AdS boundary. These solutions are to be
mapped into states in the boundary CFT, so we should associate them
with particular black hole microstates. Moreover, as emphasized in
\citep{Freivogel:2005qh}, these states will yield distinctive correlators
which lead to the hope that there might be a practical way to select
such states from a thermal ensemble. For large black holes in AdS
(and many other examples in string theory) the Bekenstein-Hawking
entropy $S_{BH}$ is identified with the logarithm of the number of
microstates. This leads to the conclusion that the number of de Sitter
bubble states should be bounded by $e^{S_{BH}}$. 

However it is not entirely clear how to interpret the Gibbons-Hawking
entropy of de Sitter space \citep{Gibbons:1977mu}. Some earlier discussions
of this in the context of AdS/CFT can be found in \citep{Lowe:2004zs}.
Another interesting attempt at interpreting de Sitter entropy in terms
of effective field theory can be found in \citep{Cohen:1998zx}. Suppose
we imagine effective field theory in a de Sitter background with an
ultraviolet energy cutoff $\Lambda$ and an infrared length cutoff
$L$. If we identify $L$ with the horizon size of our present universe
and the number of possible states in the effective field theory with
the Gibbons-Hawking entropy, we obtain $\Lambda=100$ MeV. A more
stringent bound is obtained if we restrict attention to those states
with small gravitational back-reaction, demanding that the total energy
not be inside it's own Schwarzschild radius. Again taking $L$ to
be the horizon size today, we obtain $\Lambda=10^{-2.5}$ eV. Clearly
there does not seem to be a conventional effective field theory description
that extends from horizon size scales down to scales probed in accelerator
experiments. 

In this work we advocate a variant on this idea that receives support
from some numerical coincidences. It was assumed above that the ultraviolet
cutoff is a simple uniform cutoff everywhere in spacetime. However
we know, for example from \citep{Hamilton:2005ju,Hamilton:2006az,Hamilton:2006fh},
that there is a complex relation between a UV cutoff in the bulk and
physics in the CFT on the boundary, and that this is in general state
dependent. Since we don't have a straightforward way to identify the
de Sitter bubble states in the CFT, we will make the optimistic assumption
that the bulk state is encoded using perhaps the most efficient coding
one could imagine, an adaptive stepsize with many points appearing
near aggregations of matter/energy, and few points appearing away
from such regions. The current entropy of the universe is dominated
by cosmic microwave background photons with $S_{CMB}\sim10^{85}$
\footnote{The horizon entropy of black holes formed after the big bang can easily
dominate over the entropy of CMB photons. We will choose to count
entropy by not averaging over the possible internal states of these
objects, so their contribution will remain subdominant.%
} and with Compton wavelengths of around $\lambda=1$mm. Current experiments
have tested effective field theory out to about $\Lambda=1$ TeV,
so an estimate of an upper bound on the log of number of possible
states in effective field theory would be \[
S_{tot,CMB}=(\lambda\Lambda)^{3}S_{CMB}=10^{130}.\]
 Of course it might be something of an overestimate to require so
many states associated with a single CMB photon. Taking into account
correlations between different states would certainly lower the total
number, but it is not clear how to estimate the magnitude of this
effect. Suppose instead we estimate a lower bound by considering states
associated with individual hydrogen atoms. There are approximately
$10^{80}$ atoms in our horizon volume, with a size of $\lambda=5\times10^{-11}$m
. Estimating the associated log of the number of states gives \[
S_{tot,atoms}=(\lambda\Lambda)^{3}10^{80}=10^{103}.\]
So we see these crude estimates on the number of possible states in
an adaptive stepsize effective field theory brackets the Gibbons-Hawking
entropy\[
S_{tot,atoms}<S_{GH}=10^{122}<S_{tot,CMB}\]
Henceforth we will adopt the view that the Gibbons-Hawking entropy
counts the number of possible states in effective field theory, albeit
one with a rather clever UV cutoff.

According to this picture, if we want to have a useful effective field
theory description around a semiclassical de Sitter background, the
total number of available states should saturate the Gibbons-Hawking
entropy. This will be our criterion for deciding when a local observer
would regard herself as living in a semiclassical de Sitter region.
Most states in conformal field theory correspond to backgrounds without
a geometric interpretation. Likewise one can expect many states where
some observables correspond to a particular geometric background,
but many other observables do not, causing an effective field theory
description to break down. So this criterion leads to the conclusion
that if the CFT states are to be identified with black hole microstates,
one must respect the bound \begin{equation}
S_{BH}>S_{GH}\label{eq:enbound}\end{equation}
in order that the interior of the bubble have an interpretation as
a semiclassical spacetime.

Another logical possibility is that the bubble solutions require extra
degrees of freedom, beyond those present in a single CFT, as advocated
in \citep{Freivogel:2005qh}. In that picture, the CFT on the left
boundary of Schwarzschild-AdS is regarded as being deformed. The correlations
between those degrees of freedom and those in the CFT on the right
induce a mixed state reduced density matrix, when the left boundary
CFT is traced over. These ideas were most precisely stated for solutions
such as figure \ref{fig:time symmetric}, where the asymptotic AdS
region is replaced by a segment of de Sitter space for a finite period
of time. It was argued in \citep{Freivogel:2005qh} that this corresponds
to a non-local perturbation of the left CFT. If one goes beyond the
thin-wall limit, we expect similar physics to emerge from a suitable
local irrelevant perturbation of the left CFT. In fact one could dispense
entirely with the right boundary in this picture, and study the CFT
deformation on the left corresponding to the bubble. This is closely
related to the proposal of \citep{Gubser:1998iu,Gubser:1998kv,Intriligator:1999ai,Danielsson:2000ze}
where an irrelevant perturbation of $\mathcal{N}=4$ $SU(N)$ Yang-Mills
was conjectured to be dual to gravity is asymptotically flat space.
Related ideas where the asymptotically AdS structure is changed via
CFT perturbations also appears in \citep{Hertog:2004rz,Hertog:2005hu}%
\footnote{There the perturbation was by a wrong sign marginal operator. However
this still needed a regulator to make the deformed CFT well-defined,
which can be re-expressed as adding additional irrelevant operators.%
}. If this picture was consistent, it would offer a very different
interpretation of de Sitter entropy, since there are an unlimited
number of UV degrees of freedom in the left CFT that might be associated
with bubble geometries, and hence no apparent bound by the black hole
entropy. However there is no strong evidence that effective field
theories deformed by these irrelevant perturbations really have continuum
limits, suggesting the UV completions are ill-defined. This is good
news for AdS/CFT as a nonperturbative description of string theory,
since in order for it to be complete and self-consistent, one should
not need to add additional degrees of freedom every time a quantum
fluctuation creates a region of spacetime causally disconnected from
the boundary.

\subsubsection*{Implications for de Sitter bubbles \label{sub:Implications-for-de}}

Time symmetric bubbles (and solutions related by diffeomorphism) always
respect the bound $S_{GH}>S_{BH}$, \citep{Freivogel:2005qh}. So
we conclude that when the microscopic description of the solution
is taken into account via the CFT, there are not enough available
states to represent a full semiclassical spacetime inside the bubble.
These solutions are therefore spurious once quantum effects are taken
into account. 

All solutions where the bubble wall stays in a single region of the
Schwarzschild-anti de Sitter Penrose diagram fall into this class.
Quantum tunneling solutions involving smooth initial data, as in \citep{Farhi:1989yr},
involve transitions between classical solutions of this type. We see
therefore that the rate of tunneling up the potential vanishes in
the AdS/CFT framework for quantum gravity. Another argument for the
vanishing of this tunneling rate is given in appendix \ref{sec:Detailed-balance}.

On the other hand, time asymmetric solutions of the type shown in
figure \ref{fig:Time-asymmetric-perturbed} exist where the bound
(\ref{eq:enbound}) is satisfied. These solutions pass the quantum
consistency condition described above. In classical relativity it
is not clear whether a state emerging from singular initial data is
physically meaningful. In a complete formulation of quantum gravity
such as AdS/CFT this objection must be revisited. In the case at hand
we have a set of solutions that involve an initial singularity, however
all the indications are that these correspond to well-defined states
in the CFT. We regard these black hole microstates as the most promising
way to obtain a description of inflation embedded in the AdS/CFT correspondence.

\medskip{}

\begin{acknowledgments}
I thank G.L. Alberghi and R. McNees for helpful discussions. This
research is supported in part by DOE grant DE-FG02-91ER40688-Task
A.
\end{acknowledgments}
\appendix

\section{Generic \label{sec:Appendix:-Generic-perturbations}perturbations
of bubble solutions}

Various solutions in this paper, have the property that an asymptotically
AdS region meets another asymptotic region at some instance in time.
Here we will show that such behavior is highly non-generic, and that
once perturbations are taken into account, the would-be Cauchy horizon
that emanates from this intersection point instead becomes a null
singularity, once leading order effects are considered. It is expected
this null singularity will become spacelike, once higher order effects
are incorporated.

To see this, we consider the following analog problem that is believed
to capture the essential physics. We take neutral matter in the form
of a scalar field of mass $m_{s}$ and set up the equations of motion
in a spherically symmetric ansatz. The solutions should be qualitatively
the same as matter with nontrivial angular momentum. It is helpful
to set up the Einstein equations with the variables similar to those
used in \citep{Poisson:1990eh}

\[
ds^{2}=g_{ab}dx^{a}dx^{b}+r(x^{a})^{2}d\Omega^{2}\]
 with the radius and time directions denoted by $x^{a}$, $a=1,2$.
It is helpful to define \begin{eqnarray*}
f & \equiv & g^{ab}r_{,a}r_{,b}\equiv1-2m(x^{a})/r\\
\kappa & \equiv & -m(x^{a})/r^{2}\\
T & = & T_{\, a}^{a}\,,\,\, P=T_{\,\theta}^{\theta}=T_{\,\phi}^{\phi}\end{eqnarray*}
so that the Einstein equations become

\begin{eqnarray*}
r_{;ab}+\kappa g_{ab} & = & -4\pi r(T_{ab}-g_{ab}T)\\
m_{,a} & = & 4\pi r^{2}(T_{a}^{\, b}-\delta_{a}^{\, b}T)r_{,b}\,.\end{eqnarray*}
Here the stress energy tensor is defined to include matter contributions
as well as the contribution from the cosmological constant. Conservation
of the stress energy tensor yields the equation\[
(r^{2}T^{ab})_{;b}=(r^{2})^{,a}P\,.\]
At this point we specialize to compute the behavior of the metric
near the boundary of AdS, near point P where the bubble wall reaches
the cylinder at spacelike infinity as shown in figure \ref{fig:time symmetric}.
Outside the future cone of P and exterior to the bubble, the mass
function takes the form\[
m(r)=m_{0}+\Lambda r^{3}/6\]
with the cosmological constant related to the AdS radius of curvature
$R$ by $\Lambda=-3/R^{2}$. The main point is that if the geometry
is no longer asymptotically AdS to the past of some point, generic
matter perturbations will induce both the normalizable and non-normalizable
modes with respect to asymptotically AdS geometry. Both of these perturbations
become normalizable in the geometry to the past of point P. The stress
energy of the {}``non-normalizable'' modes diverges as $r\to\infty$
for $\Delta>3$\begin{eqnarray*}
T_{tt} & \sim & r^{2(\Delta-3)}\\
T_{rr} & \sim & r^{2(\Delta-4)}\\
T_{\theta\theta} & \sim & T_{\phi\phi}\sim r^{2(\Delta-3)}\,.\end{eqnarray*}
Here the conformal dimension $\Delta$ is related to the mass of the
scalar field $m_{s}$ by $\Delta=3/2+\sqrt{9/4+m_{s}^{2}R^{2}}$.
For sufficiently large $\Delta$ this will dominate over the cosmological
constant contribution as $r\to\infty$. In this limit, one obtains
the equation\begin{equation}
\Box m=-16\pi^{2}r^{3}T_{ab}T^{ab}\,.\label{eq:mlap}\end{equation}
 Our strategy will be similar to that of \citep{Poisson:1990eh,Bonanno:1994ma},
namely we will treat the back-reaction of the stress energy on the
geometry at leading order in perturbation theory around an asymptotically
AdS geometry. This will lead to a curvature singularity as one approaches
the future light-cone of the point P. 

To proceed, we solve (\ref{eq:mlap}) using Kruskal coordinates. Outside
the future light-cone of point P, we can use the pure AdS metric near
infinity\[
ds^{2}=(1-\Lambda r^{2}/3)dudv+r^{2}d\Omega^{2}\]
where the new coordinates are related to the old coordinates by\begin{eqnarray*}
u=-t+r^{*} & , & v=t+r^{*}\\
r^{*} & = & \sqrt{\frac{3}{\left|\Lambda\right|}}\arctan\left(\sqrt{\frac{\left|\Lambda\right|}{3}}r\right)\,.\end{eqnarray*}
In these coordinates (\ref{eq:mlap}) takes a simple form \[
\frac{\partial^{2}m}{\partial u\partial v}\sim-cr^{4\Delta-9}\]
where $c$ is a constant dependent on the amplitude of the matter
perturbation. The solution for $m$ is divergent as one approaches
the future light-cone of the point P ($v=\frac{\pi}{2}\sqrt{\frac{\left|\Lambda\right|}{3}}$)\[
m\sim\frac{1}{(v-\frac{\pi}{2}\sqrt{\frac{\left|\Lambda\right|}{3}})^{4\Delta-7}}\,.\]

This provides strong evidence for at least a null singularity extending
out on the future light-cone of the point P. Eventually higher order
terms will come to dominate and the perturbative analysis breaks down,
as is expected in \citep{Poisson:1990eh,Bonanno:1994ma}. It is expected
that the null singularity becomes a true spacelike singularity in
the full non-linear analysis.

\section{\label{sec:Detailed-balance}Detailed balance}

The implications of detailed balance for quantum tunneling from flat
space to de Sitter space has been previously studied in \citep{guth}.
Let us reconsider that argument in the present context. Detailed balance
implies the transition probabilities between any two states $i,j$
are related by\[
\Gamma_{ij}p_{i}=\Gamma_{ji}p_{j}\]
where $\Gamma_{ij}$ is the transition rate from state $i$ to state
$j$ and $p_{i}$ is the equilibrium probability of state $i$. Consider
$AdS_{4}$ in thermal equilibrium, with large radius of curvature
$R$ and temperature $T$, in the regime where Schwarzschild-anti-de
Sitter space dominates the canonical ensemble. Let us suppose there
exist some states in this ensemble corresponding to a small region
of the Schwarzschild-AdS spacetime tunneling off into a de Sitter
bubble. We assume the initial size of the bubble is the smallest length
scale in the problem. Applying detailed balance to a transition between
two such states, we obtain

\begin{equation}
\Gamma_{up}=\Gamma_{CDL}e^{S_{GH}-S_{BH}}\thicksim e^{-aR^{2}T^{2}}\label{eq:rans}\end{equation}
where $\Gamma_{CDL}$ is the Coleman-De Luccia tunneling rate, $\Gamma_{up}$
is the rate of tunneling up the potential, $S_{BH}$ is the entropy
of the Schwarzschild-AdS spacetime and $S_{GH}$ is the Gibbons-Hawking
entropy of a single de Sitter bubble, and $a$ is a coefficient independent
of $T$ \citep{Klebanov:1996un}. In this limit $S_{BH}\sim aR^{2}T^{2}$
which we take to be much larger than $S_{GH}$. We expect $\Gamma_{CDL}$
to be independent of $R$ and $T$ in this limit, since it should
be dominated by local physics.

The temperature dependence of the upward tunneling rate can be estimated
as follows. Suppose an excited state with energy $M$ is assembled,
capable of tunneling to the de Sitter bubble. The probability of such
a state will be $e^{-M/T-S_{BH}}$. The rate of upward transitions
will then be\begin{equation}
\Gamma_{up}=\Gamma_{tunnel}e^{-M/T}\label{eq:wans}\end{equation}
where $\Gamma_{tunnel}$ and $M$ should be independent of $T$ and
$R$ by locality. However the two rates (\ref{eq:rans}) and (\ref{eq:wans})
have different temperature dependencies, so cannot match. This implies
the rate of quantum tunneling $\Gamma_{tunnel}$ must vanish. This
provides an argument, independent to that given in section \ref{sub:Implications-for-de},
that quantum tunneling up the potential, as envisioned in \citep{Farhi:1989yr},
does not happen. Nevertheless we do expect transitions between states
of the type shown in figure \ref{fig:Time-asymmetric-perturbed} and
Schwarzschild-AdS geometries, and these should be governed by the
rate (\ref{eq:rans}).

\bibliographystyle{brownphys}
\clearpage\addcontentsline{toc}{chapter}{\bibname}\bibliography{bubble}

\end{document}